\PassOptionsToPackage{table,xcdraw,svgnames}{xcolor}
\documentclass{article} % For LaTeX2e
\usepackage{paper,times}

\usepackage{amsmath,amsfonts,bm}

\newcommand{\name}{Alpha${^2}$}
\def\eqref#1{equation~\ref{#1}}
\def\1{\bm{1}}

\DeclareMathAlphabet{\mathsfit}{\encodingdefault}{\sfdefault}{m}{sl}
\SetMathAlphabet{\mathsfit}{bold}{\encodingdefault}{\sfdefault}{bx}{n}

\usepackage{xcolor}
\usepackage{url}
\usepackage{graphicx}
\usepackage[lofdepth,lotdepth]{subfig}
\usepackage{booktabs}
\usepackage{varwidth}
\usepackage{wrapfig}
\usepackage{xcolor} 
\usepackage{multicol}
\usepackage{authblk}
\usepackage{multirow}
\usepackage{enumitem}
\usepackage[bottom,stable]{footmisc}
\usepackage{hyperref}
\makeatletter
\newcommand*{\addFileDependency}[1]{
  \typeout{(#1)}
  \@addtofilelist{#1}
  \IfFileExists{#1}{}{\typeout{No file #1.}}
}
\makeatother

\title{\name: Discovering Logical Formulaic Alphas using Deep Reinforcement Learning}

\author[ ]{\centering
{\large
Feng Xu$^{1,2*}$, Yan Yin$^*$, Xinyu Zhang$^{1,2}$, Tianyuan Liu$^{1,2}$, \\
Shengyi Jiang$^3$, and Zongzhang Zhang$^{1,2\dagger}$}
}
\affil[1]{National Key Laboratory for Novel Software Technology, Nanjing University, Nanjing, China}
\affil[2]{School of Artificial Intelligence, Nanjing University, Nanjing, China}
\affil[3]{The University of Hong Kong, Hong Kong, China}
\affil[ ]{\texttt{xufeng@lamda.nju.edu.cn}, \texttt{yy855@cornell.edu}, \newline\texttt{\{zhangxinyu, liutianyuan\}@lamda.nju.edu.cn}, \texttt{syjiang@cs.hku.hk}, \texttt{zzzhang@nju.edu.cn}}

\begin{document}

\maketitle
\begin{abstract}
\let\thefootnote\relax\footnotetext{$^*$Equal Contribution\ $^\dagger$Corresponding Author}
Alphas are pivotal in providing signals for quantitative trading. The industry highly values the discovery of formulaic alphas for their interpretability and ease of analysis, compared with the expressive yet overfitting-prone black-box alphas. In this work, we focus on discovering formulaic alphas. Prior studies on automatically generating a collection of formulaic alphas were mostly based on genetic programming (GP), which is known to suffer from the problems of being sensitive to the initial population, converting to local optima, and slow computation speed. Recent efforts employing deep reinforcement learning (DRL) for alpha discovery have not fully addressed key practical considerations such as alpha correlations and validity, which are crucial for their effectiveness. In this work, we propose a novel framework for alpha discovery using DRL by formulating the alpha discovery process as program construction. Our agent, \name, assembles an alpha program optimized for an evaluation metric. A search algorithm guided by DRL navigates through the search space based on value estimates for potential alpha outcomes. The evaluation metric encourages both the performance and the diversity of alphas for a better final trading strategy. Our formulation of searching alphas also brings the advantage of pre-calculation dimensional analysis, ensuring the logical soundness of alphas, and pruning the vast search space to a large extent.  Empirical experiments on real-world stock markets demonstrates \name's capability to identify a diverse set of logical and effective alphas, which significantly improves the performance of the final trading strategy. The code of our method is available at \href{https://github.com/x35f/alpha2}{https://github.com/x35f/alpha2}.

\end{abstract}

\section{Introduction}
In quantitative investment, alphas play a pivotal role in providing trading signals. Serving as the foundation for strategic decision-making, alphas transform raw market data, such as opening and closing prices, into actionable signals such as return predictions. These signals inform traders' decisions and shape their strategies. Uncovering high-performance alphas that can withstand market fluctuations has long been a focal point in financial research.

%introducing formulaic alphas and machine learning-based alphas
Alphas are broadly categorized into two groups: formulaic alphas and black-box alphas. Formulaic alphas,  in the form of operators and operands, are widely adopted because of their straightforward mathematical expressions and ease of analysis. They encapsulate market dynamics into succinct formulas. A textbook example is $\frac{close-open}{high-low}$, a mean-reversion alpha. Conversely, black-box alphas leverage advanced machine learning algorithms, such as deep learning and tree-based models ~\citep{ke2017lightgbm, chen2016xgboost}, to directly transform a group of inputs to a numerical signal. 
Despite their high expressivity and handy end-to-end nature, they come with their own set of challenges. The lifetimes can be notably shorter, and training these models demands careful tuning of hyper-parameters. A widely held opinion is that formulaic alphas, given their simplicity and transparency, exhibit resilience to market fluctuations and are often more enduring than the machine-learning counterparts. Our work focuses on the discovery of formulaic alphas.

% genetic programming and reinforcement learning
Traditionally, the discovery of formulaic alphas is often attributed to the intuition and insights of a trader, usually grounded in economic fundamentals. However, as modern computational capabilities advance, algorithmic techniques are also employed to find formulaic alphas~\citep{AlphaGen,cui2021alphaevolve,zhang2020autoalpha}. These methods can identify alphas that satisfy specific criteria without the need for constant human oversight.
Genetic Programming (GP)~\citep{koza1994genetic} based solutions, such as those detailed in ~\citep{zhang2020autoalpha}, have gained traction as popular tools for alpha discovery. These solutions maintain a population of expressions that undergo stochastic modifications, such as crossover and mutations. AlphaGen~\citep{AlphaGen} pioneers the usage of Reinforcement Learning (RL) to discover alphas, which leverages an RL agent to sequentially produce alphas. While their results showcased the potential of RL in this domain, their adoption of the techniques can be improved. These existing works exhibit two weaknesses that hinder their practical use. First, they are not able to find formulaic alphas built from more primitive operators or deeper structures. This problem is even worse for GP-based methods because of their sensitivity to initial population distributions and high computational demands. Second, existing methods tend to use the performance of alpha as the only evaluation metric, producing alphas with high information correlation (IC) and low interpretability. 

From the view of practical strategies for real-market data, alphas should satisfy two properties. First, as outlined in ~\citep{Tulchinsky2015FindingAA}, diversity among alphas plays a important role in constructing robust trading strategies. This diversity helps mitigate the risk of overfitting, ensuring that strategies remain resilient when facing market volatility.  Second, alphas should be logically sound according to certain rules, such as dimension consistency. For example, performing an addition between the open price and volume should be avoided, since they are not of the same dimension. GP-based methods directly modify the structure of expressions, while AlphaGen constructs an expression in the form of Reverse Polish Notation, token by token. Both methods can only perform dimensional analysis after an alpha is fully constructed. Being unable to prune the search space in advance, a lot of computational efforts are wasted.

% search space
One key challenge of alpha discovery lies in its large search space. To illustrate, consider a task involving 40 binary operators and 20 operands. For an alpha constituted of up to 15 operators, the search space swells to an overwhelming size of approximately $10^{63}$. Performing brute-force search on this space is impractical. In the AlphaGo class of algorithms~\citep{alphadev, alphago, alphazero}, RL-guided Monte Carlo Tree Search (MCTS) has demonstrated strong ability in finding solutions in large search spaces, such as Go, Chess, Shogi, Starcraft, and assembly programs.

%our method
To address the challenges observed in previously discussed frameworks and to discover alphas for practical use, we present a novel alpha discovery approach that combines RL with MCTS to generate alphas that are logical and less correlated. Drawing inspiration from AlphaDev~\citep{alphadev}, we conceptualize an alpha as a program, akin to an assembly program, assembled incrementally. Such programs can be seamlessly translated into expression trees to be calculated. Meanwhile, such construction of an alpha can easily prune the search space in advance according to predefined rules. We then encapsulate this generation process within an environment. Subsequently, by leveraging a refined value estimation and policy guidance from DRL, we efficiently focus the search for diverse, robust, and high-performance alphas.
Empirical studies validate the efficacy of our framework, confirming that alphas searched via our method surpass those discovered through traditional methods, in terms of its performance, correlation, and validity.

The primary contributions of our work are:
\begin{itemize}[leftmargin=10pt]
\item We reconceptualize the task of generating formulaic alphas as a program generation process. The assembly of the alpha program enjoys the benefits of pruning the search space to a large extent. 
\item We present a novel search algorithm for formulaic alpha generation, utilizing the strength of DRL.
\item Our experimental results validate the efficacy of our approach. We achieve a substantial reduction in search space and demonstrate the capability to discover logical, diverse, and effective alphas.
\end{itemize}

\section{Related Works}
\textbf{Symbolic Regression}:
Symbolic Regression (SR) is a machine learning technique aiming to discover mathematical expressions to fit a dataset. The search for alphas can be seen as a way for SR to predict the market's return, which is highly correlated with our problem setting. However, the data in the financial market typically has a low signal-to-noise ratio, making accurate predictions from expressions of operators and operands impossible. Techniques like genetic programming, Monte Carlo Tree Search, and neural networks have been applied to symbolic regression. \cite{mundhenk2021symbolic} introduce a hybrid neural-guided GP approach, utilizing the power of RL to seed the GP population. \cite{sahoo2018learning} use a shallow neural network structured by symbolic operators to identify underlying equations and extrapolate to unseen domains. \cite{kamienny2023deep} propose a MCTS-based method, using a context-aware neural mutation model to find expressions. 

\textbf{Auto Generation of Formulaic Alphas}:
Formulaic alphas provide interpretable trading signals based on mathematical expressions involving market features.
Automated discovery of alphas has gained traction in quantitative trading. Genetic Programming has been widely applied to find trading strategies by evolving mathematical expressions. 
AutoAlpha~\citep{zhang2020autoalpha} uses Principal Component Analysis to navigate the search path from existing alphas. 
AlphaEvolve~\citep{cui2021alphaevolve} utilizes AutoML techniques to a new class of alphas that are different from previous classes of formulas and machine learning models. 
\cite{AlphaGen} first propose to use RL to generate formulaic alphas in the form of Reverse Polish Notation, and introduce a framework to automatically maintain the best group of alphas. Although the result of AlphaGen shows great improvement over previous methods, their framework of the RL task can be further improved. Due to the sparse nature of alpha search space, the Markov Decision Process defined in AlphaGen leads to highly volatile value estimations, and the policy generates similar alpha expressions.

\section{Problem Formulation}

\subsection{Definition of Alpha}

In this study, we focus on finding a day-frequency trading strategy in a stock market consisting of \(n\) distinct stocks spanning \(T\) trading days. For a stock dataset consists of $D$ trading days, on every trading day \(d \in \{1,2,...,D\}\), each stock $i$ is represented by a feature vector \(x_{d,i} \in \mathbb{R}^{m\tau}\). This vector encapsulates \(m\) raw features, including open, close, high prices, etc, in the past $\tau$ days.  $\tau$ is decided according to the expression of the alpha and availability of data. An alpha is a function $\zeta$ that transforms the features of a stock into a value \(z_{d,i} = \zeta(x_{d,i}) \in \mathbb{R}\). These values of alphas are subsequently utilized in a combination model to form the trading signals. In the rest part of the paper, we omit the date index $i$ for brevity and operate on the $D$-day stock dataset.

\subsection{Evaluation Metrics for an Alpha}
To assess an alpha's efficacy, the primary metric used is the Information Correlation (IC), which is computed as the average Pearson Correlation Coefficient between the alpha value $z$ and market return $\mu$ in $D$ days. It is mathematically expressed as:
\begin{equation} 
\mathrm{IC}(z, \mu) =  \frac{\sum_{d=1}^D\frac{\mathrm{Cov}(z_d,\mu_d)}{\sigma_{z_d}\sigma_{\mu_d}}}{D}, \label{ic_definition}
\end{equation}
where $\mathrm{Cov}$ computes the covariance and $\sigma$ computes the standard deviation.

Additional metrics used to further evaluate alphas include Rank IC, Max Draw Down (MDD), turnover (TVR), and Sharpe Ratio. While these metrics are not the primary targets of optimization within our framework, they hold substantial importance in practical trading scenarios and provide a comprehensive understanding of alpha performance, and our framework can be easily customized to these evaluation metrics.

\section{Methodology}
\subsection{Alpha Discovery as Program Generation}
Formulaic alphas are structured compositions of operators and operands. Drawing inspiration from AlphaDev's approach to algorithm generation, we reconceptualize the task of alpha discovery as constructing an ``alpha program".
\subsubsection{Operators, Operands and Instructions}
\begin{table}[t]
\vspace{-10mm}
\centering
\subfloat[Operators]{
\begin{tabular}{c|c}
    \toprule
    \textbf{Category}  &  \textbf{Examples}\\
    \hline
    Unary & Abs, Ln, Sign, ...\\
    Binary  & Add, Sub, Mul, TS-Mean, ...\\
    Ternary & Correlation, Covariance, ...\\
    Indicator & Start, End\\
      \bottomrule
\end{tabular}}
\quad
\subfloat[Operands]{
\begin{tabular}{c|c}
    \toprule
 \textbf{Category}  &  \textbf{Examples}\\
    \hline
    Scalar & 0, 0.1, 0.5, 1, 3, 5, 15, ...\\
    Matrix & open, close, high, low, vwap, ...\\
    Register & Reg0, Reg1, ...\\
    Placeholder & Null \\
      \bottomrule
\end{tabular}}
\caption{Operators and operands}
\label{tab:ops}
\end{table}

An alpha program is built from a series of instructions, where each instruction is characterized as a 4-element tuple $(Operator,\ Operand1,\ Operand2,\ Operand3)$. Operators are grouped into unary, binary, ternary, and indicator types based on the type and number of operands they engage. The indicator operators mark the start and end of an alpha program. Operand types include scalar values, matrix data, register storage, and a placeholder. Scalar operands are parameters for operators. Matrix operands are input features of the market, such as $open$ and $close$. Registers are used for storing intermediate results. The placeholder operand, $Null$, is used to align the instructions to a 4-element tuple. Examples of operators and operands are provided in Tab.~\ref{tab:ops}.
\subsubsection{Translating an alpha program into a computation tree}
\begin{table}[t]
  \begin{varwidth}[b]{0.6\linewidth}
    \centering
    \begin{tabular}{c|c|c|c|c}
        \toprule
        Operator & Operand1 & Operand2 & Operand3 & Register\\
        \hline
        Start & Null & Null & Null & \\
        \textcolor{blue}{Sub} & \textcolor{blue}{close} &  \textcolor{blue}{open} &  \textcolor{blue}{Null}&  \textcolor{blue}{Reg0}\\
         \textcolor{green}{Sub} & \textcolor{green}{high} & \textcolor{green}{low} & \textcolor{green}{Null} & \textcolor{green}{Reg1}\\
        \textcolor{orange}{Div} & \textcolor{orange}{Reg0} & \textcolor{orange}{Reg1} & \textcolor{orange}{Null} & \textcolor{orange}{Reg0}\\
        End & Null & Null& Null\\
              \bottomrule
    \end{tabular}
    \caption{An example alpha program of $\frac{close -open}{high-low}$}
    \label{tab:alpha_program}
  \end{varwidth}
  \hfill
  \begin{minipage}[b]{0.35\linewidth}
    \centering
    \includegraphics[width=0.7\linewidth]{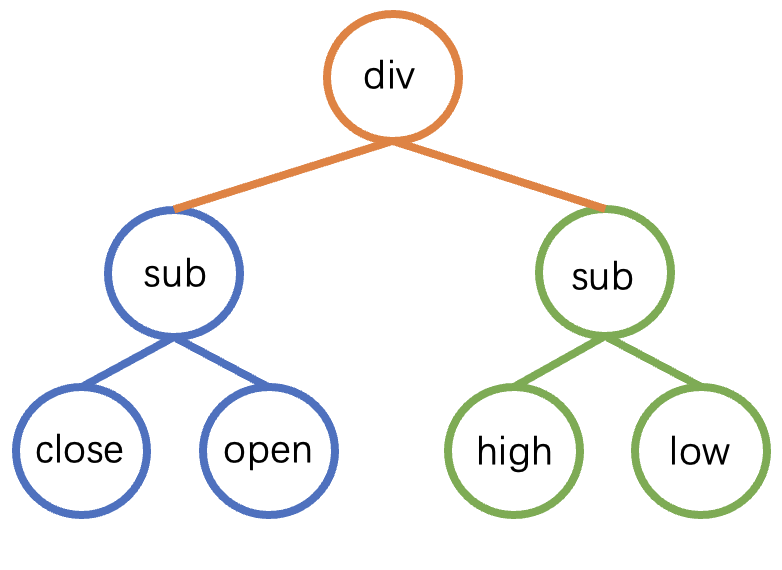}
    \captionof{figure}{Expression tree}
    \label{fig:alpha_program}
  \end{minipage}
\end{table}

To actually compute an alpha program, we need to convert the program into formats that the computer can understand. A computational tree is built from alpha instructions in a bottom-up way. Tab.~\ref{tab:alpha_program} and Fig.~\ref{fig:alpha_program} provide an example of this transformation. Programs begin with the instruction tuple $(Start,\ Null,\ Null,\ Null)$. Then, instructions are translated one by one to build an expression tree. The color coding in the table corresponds to the colored nodes of the expression tree. Each colored node in the tree is the result of the execution of its matching colored instruction in the alpha program. The instructions marked by blue and green fill two registers. Note that the register assignment is implicit, if an instruction doesn't utilize registers, its output is stored in the first available register. Then, the instruction marked by orange performs a division between the values in the two registers. For an instruction employing a single register, the output replaces the current value. When an instruction involves two registers, the computed result replaces the Reg0 value while Reg1 is emptied. The  $(End,\ Null,\ Null,\ Null)$ instruction marks the termination of an alpha program. Evaluating an alpha program involves reading values from the Reg0 register either during program construction or after its completion. This implicit approach to register management has been proven effective in our experiments.

\subsubsection{The MDP for Reinforcement Learning Task}
Given the established operators, operands, and instructions, we can construct a task suitable for RL. This task is defined as a Markov decision process (MDP), denoted as $(\mathcal S, \mathcal A, p, r, \gamma, \rho_0)$, where $\mathcal S$ is the state space, $\mathcal A$ is the action space, $p(\cdot|s,a)$ is the transition probability, $r(s,a)\in[0,R_{\max}]$ is the reward function, $\gamma\in(0,1)$ is the discount factor, and $\rho_0(s)$ is the initial state distribution. 

In our alpha program environment, the state space, $\mathcal S$, contains all potential alpha programs. Each state $s$ corresponds to a unique alpha function $\zeta$. A state $s$ is a vectorized representation of the alpha function $\zeta$.
The action space, $\mathcal A$, is the set of all possible instructions.
The transition probability, $p(\cdot|s,a)$, is deterministic, which takes the values 1 for the next alpha program build after applying an action. 
The reward, denoted as $r(s_t, a_t, s_{t+1})$, is determined by the increase in the evaluation metric when arriving at $s_{t+1}$ after applying action $a_t$. 
The evaluation metric function is denoted as $\mathrm{Perf}(\zeta)$, which takes the alpha expression of the state as input. Since our transition is deterministic, the reward is computed as $r(s_t, a_t, s_{t+1}) = \mathrm{Perf}(\zeta_{t+1}) - \mathrm{Perf}(\zeta_t)$. The definition of the evaluation metric is primarily IC, but we have refined it, which we will detail later.
The discount factor, $\gamma$, is a hyper-parameter to control the length of the alpha program.
The initial state distribution, $\rho_0(s)$, invariably starts from an empty program, i.e., its value is 1 for an empty program, and 0 otherwise.

\subsection{Discovering Alphas using RL}

\name\ uses a DRL agent to explore the alpha program generation task. The RL algorithm of \name\ is similar to that of AlphaDev~\citep{alphadev}, which is a modification of the AlphaZero agent~\citep{alphazero}. DRL guides a MCTS procedure using a deep neural network. The deep neural network takes the current state $s_t$, which is a vectorized representation of the alpha program $\zeta_t$, as input, and outputs action distributions and value predictions. The action distribution predicts the prior probability that an agent should take for each action, and the value predictions predict the cumulative reward that the agent should expect from the current state $s_t$.
Since the Alpha series of works has detailed the MCTS process, we do not elaborate on it in this paper. The next paragraphs focus on key improvements that make the search algorithm better for discovering formulaic alphas.

\subsection{Discovering robust, diverse and logical alphas}

\subsubsection{Discovering robust alphas}
Our approach to estimating the value of child nodes introduces a nuanced deviation from conventional methodologies. In traditional MCTS, the mean operator is often used to calculate the values of child nodes.  Our empirical findings indicate that this operator falls short during the initial phases of the algorithm, a phenomenon we attribute to the inherent sparsity of formulaic alphas. In the early tree search stage, most alphas yield non-informative signals, leading to arbitrary policy directions. This scenario calls for the adoption of a max operator, which is more adept at navigating the sparse landscape of formulaic alphas. However, simply using the max operator can lead to the discovery of parameter-sensitive alphas, which is not desired. Supporting our observation, ~\cite{dam2019generalized} state that using the mean operator leads to an underestimation of the optimal value, slowing down the learning, while the maximum operator leads to overestimation. They propose a power mean operator that computes the value between the average value and the maximum one. In our work, we take a simpler form, and leave the balance between the maximum operator and the mean operator controlled by a hyperparameter. The value estimation for a child node is formulated as 
$$Q(s,a) = r(s,a) + \beta\cdot \mathrm{mean}(V_s) + (1-\beta)\max(V_s),$$
where $\beta\in[0,1]$ is a hyperparameter controlling the balance between mean and max, $V_s$ is the value backup of a node on state $s$. Also, to further increase the validity of value estimation, especially for the mean operator, the value backup is calculated from the top-k values added to the node. That is, $V_s = \{v_1, ..., v_k\}$, where the $k$ values are stored in a min heap, so that the computation complexity is $O(\log k)$ for each new value. The RL agent, in the simulation phase, operates based on maximizing this $Q$-value.
This refined definition of $Q$-value is expected to help discover alphas that are both effective and robust to parameters.

\subsubsection{Discovering diverse alphas}
As outlined in \citep{Tulchinsky2015FindingAA}, diversity among alphas helps buliding robust trading strategies. In our framework, we incorporate such a target within the evaluation function. The diversity can be quantified by computing the correlation between alphas. For an alpha function $\zeta_t$ to be evaluated, we first compute its alpha value $z_t$ on all stocks and trading days. Then, for an already discovered set of alpha values $G=\{z^1, z^2, ..., z^n\}$, where $n$ is the number of alphas. We compute the max correlation with the current alpha value, i.e., $\mathrm{MaxCorr}(z_t, G)=\max_i \mathrm{IC}(z_t, z^i)$.
The evaluation metric is discounted according to the max Pearson Correlation Coefficient between the  alpha value and the mined alpha set: 
$$\mathrm{Perf}(\zeta_t) = (1-\mathrm{MaxCorr}(z_t, G))\cdot \mathrm{IC}(z_t, \mu),$$
This evaluation metric function encourages the discovery of low-correlation alphas by assigning higher value to alphas with low correlation with the mined alpha set, and discourages the discovery of highly correlated alphas by reducing the value of alphas with high correlation. In this way, \name\ can continuously discover diverse alphas.

\subsubsection{Ensuring the Dimensional Consistency of Alphas}

\begin{figure}[t]
    \vspace{-15mm}
    \centering
    \includegraphics[width=0.95\linewidth, trim=0 20 0 70,clip]{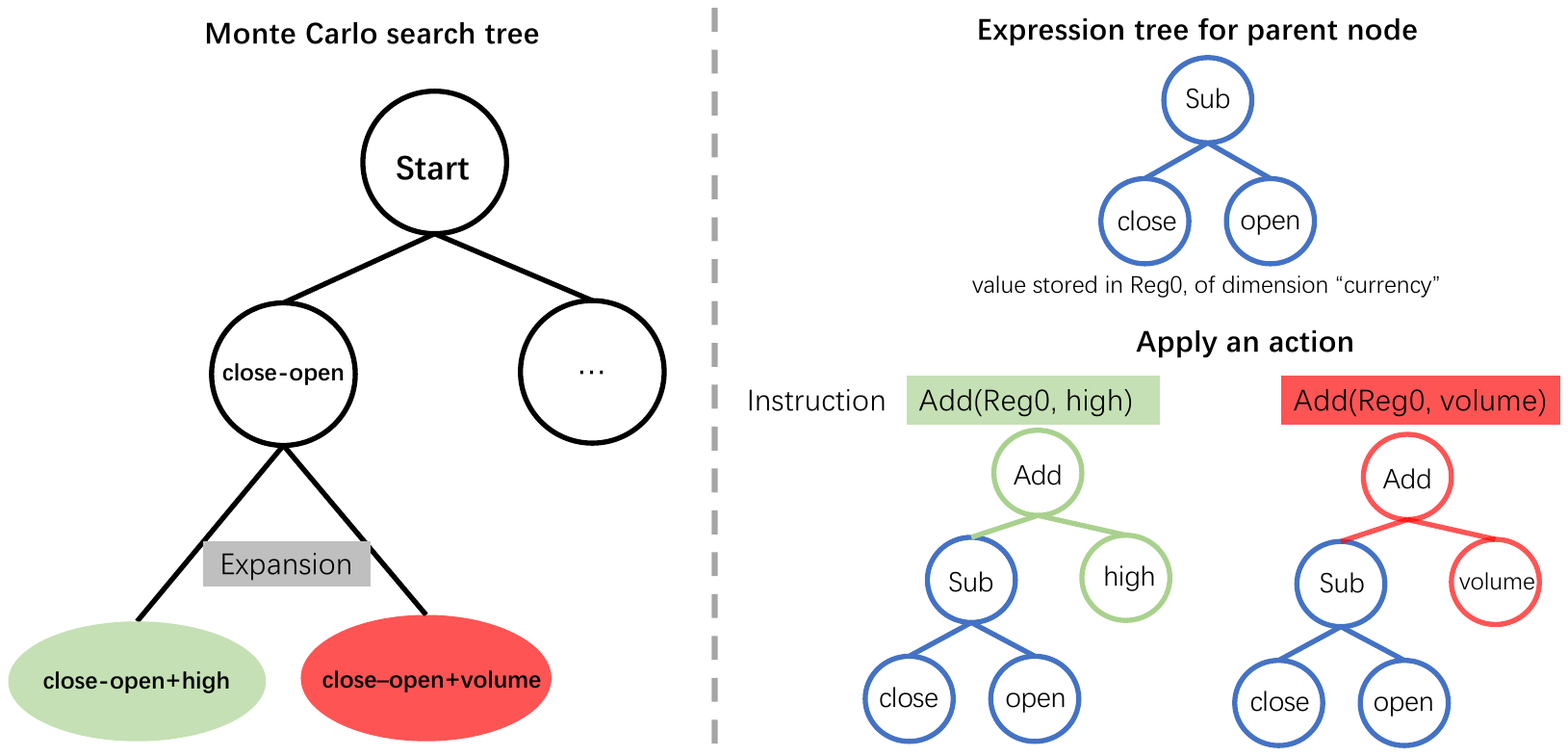}
    \vspace{-12mm}
    \caption{Example of pruning the search space according to dimensional consistency. The left part shows an ongoing search process of MCTS. The right part shows the expression trees of the corresponding alpha expressions. For an alpha expression $close - open$, which is of dimension $currency$, consider adding $high$ or $volume$ to the expression. Adding $high$ is allowed since it is of the same dimension $currency$, while adding $volume$ is forbidden because it is of another dimension. The nodes marked in green illustrates the addition of $high$, and the nodes marked in red illustrates the addition of $volume$.}
    \label{fig:dimension}
\end{figure}

\begin{wraptable}[14]{R}{0.35\textwidth}
    \centering
    \begin{tabular}{c|c}
        \toprule
        Feature & Dimension \\
        \hline
        open & currency \\
        close & currency \\
        high & currency \\
        low & currency \\
        vwap & currency \\
        volume & unit \\
        \bottomrule
    \end{tabular}
    \caption{Dimension of features}
    \label{tab:dimension}
\end{wraptable}

In real-world applications, especially in financial contexts, meaningful interactions between features are vital. Combining disparate features can lead to spurious relationships, making trading strategies derived from them unreliable. 
In the SR field, the dimension of individual input features is generally overlooked. For SR approaches in deep learning, normalization of features typically occurs during the pre-processing phase. Yet, for alpha discovery, where data is extended over both time and the count of assets, integrating normalization within the search process is preferable. This is from the fact that normalization inherently alters the data's semantics. While AlphaGen and GP-based methods have produced alphas with impressive statistical metrics, these often lack an underlying logical rationale. A defining quality of a coherent alpha expression is the dimensional consistency among its operators and operands. For instance, summing up variables like price and trade volume is fundamentally wrong due to their divergent distributions and different dimensions. Tab.~\ref{tab:dimension} lists the dimensions of the basic input features. Note that the dimension changes as the expression gets complicated. Our approach innovates by imposing rules that constrict the search space right from the node expansion stage, before actually evaluating the alpha, which is a feature not achievable in preceding methods. We maintain a record of an expression's dimension within each register, allowing for a preemptive filtration of nodes based on the dimensional requisites specified by the operators when expanding an MCTS tree. Our method to construct an alpha ensures that once a segment of the alpha's expression tree passes the dimensional check, it does not need to be reassessed.  An example of the dimension system is illustrated in Fig.~\ref{fig:dimension}.

 Traditional methods, including AlphaGen and GP-based paradigms, due to their structural limitations, are unable to prune the search space in advance. AlphaGen incrementally constructs alphas, token by token. Coupled with its use of Reverse Polish Notation for expressions, pinpointing a token's exact location within the final expression tree is ambiguous, which does not support the search space pruning. GP-based methods, with their mutation operations, remain unaware of the overall structure of expression trees, and can even perform cyclic modifications. Thus, performing dimension check is not achievable before the alpha is generated.  Employing these dimension restrictions results in a large reduction in nodes at every level compared to the unrestrained counterpart. By reducing the complexity and potential combinations in the search space, we can focus on discovering alphas that are logical. This can also reduce the chances of overfitting, which is a significant concern in quantitative finance.

 \subsection{Pipeline to generate a trading strategy}
\begin{figure}[t]
    \vspace{-10mm}
    \centering
    \includegraphics[width=\linewidth, trim=70 150 70 150,clip]{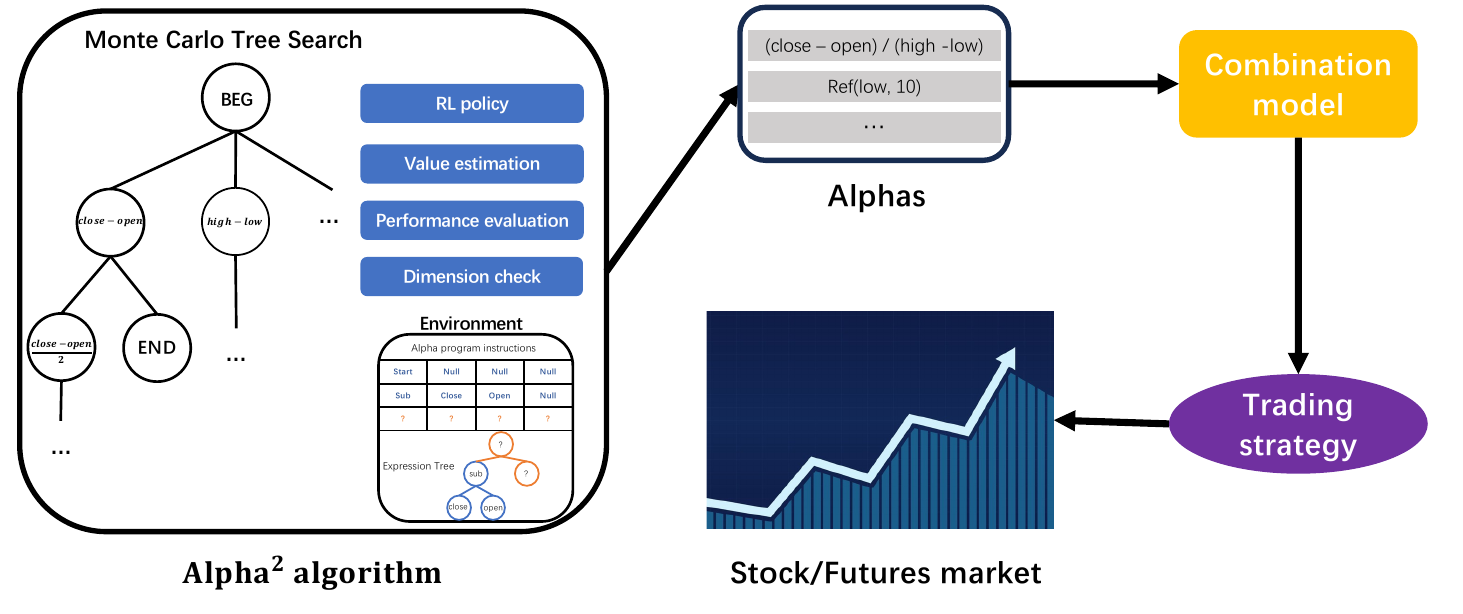}
    \caption{The pipeline for the generation of strategy}
    \label{fig:framework}
    \vspace{-5mm}
\end{figure}
 Our method focuses on generating alphas and does not provide an end-to-end solution. \name\ first produces alphas using RL-guided MCTS with the refined value estimation, performance evaluation, and dimension check. Then, a combination model takes the alphas as input and generates a trading strategy. The combination model can be customized to meet user demand, e.g., linear regression, deep neural networks, and gradient boosting trees. Fig.~\ref{fig:framework} shows the pipeline for a practical adoption of our method.

\vspace{-5mm}
\section{Experiments}
In the experiment section, we aim to demonstrate the efficacy of our method compared to existing approaches. Our primary focus is to answer the three questions:
\begin{itemize}[leftmargin=10pt]
    \item Can \name\ generate diverse and good alphas?
    \item Can alphas mined by \name\ perform better than previous methods?
    \item How do alphas mined by \name\ perform in the real-world market?
\end{itemize}
\subsection{Experiment Setup}

\textbf{Data:}. The data is acquired from the Chinese A-shares market through baostock\footnote{https://www.baostock.com}. Six raw features are selected to generate the alphas: \{open, close, high, low, volume, vwap\}. The target of our method is to find alphas that have high IC with the 20-day return of the stocks. The dataset is split into a training set (2009/01/01 to 2018/12/31), a validation set (2019/01/01 to 2020/12/31), and a test set (2021/01/01 to 2023/12/31). We use the constituents of the CSI300 and CSI500 indices of China A-shares as the stock set.

\textbf{Baselines:}  Our method is compared to several machine learning models. \textbf{MLP} uses a fully connected neural network to process the input data to strategy signals. \textbf{XGBoost} and \textbf{LightGBM} are gradient boosting frameworks. \textbf{AlphaGen} and \textbf{gplearn}~\footnote{https://github.com/trevorstephens/gplearn} are representative methods for generating a collection of alphas. We follow the open source implementations of AlphaGen~\footnote{https://github.com/RL-MLDM/AlphaGen} and Qlib~\footnote{https://github.com/microsoft/qlib}~\citep{yang2020qlib} to produce the results.

\textbf{Alpha Combination:} Our method, \name, only considers the problem of generating alphas. We use the XGBoost as the combination model. The XGBoost model is trained to fit the alpha signals to the return signals on the training dataset, using the top 20 generated alphas ranked by IC. Then, the trained model is fixed and used to predict the test dataset.

\textbf{Evaluation Metric:}
Two metrics, IC and Rank IC, are used to measure the performance of the models. The definition of IC is given in Eq.~\ref{ic_definition}. The rank information coefficient (Rank IC) measures the correlation between the ranks of  alpha values and the ranks of future returns. It is defined as the Spearman's Correlation Coefficient between the ranks of the alpha values and the future return, $\rho(z_d,r_{d})=\mathrm{IC}(\mathrm{rk}(z_d), \mathrm{rk}(r_d))$, where $\mathrm{rk}(\cdot)$ is the ranking operator.

\textbf{Code:} For an efficient and accelerated experimental process, our implementation is based on the pseudo-code that AlphaDev provides~\footnote{https://github.com/google-deepmind/alphadev}, with computational aspects handled by Jax. Experiments are run on a single machine with an Intel Core 13900K CPU and 2 Nvidia A5000 GPUs.

\subsection{IC and Correlation of Generated Alphas}

\renewcommand{\arraystretch}{1.2}
\begin{table}[htbp]
\centering
\begin{tabular}{c|cc}
\toprule
Method & IC & Correlation \\
\hline
gplearn & 0.0164$\pm$0.0167& 0.7029$\pm$0.1824\\
AlphaGen & 0.0257$\pm$0.0153& 0.3762$\pm$0.6755\\
\hline
\textbf{Ours} & \textbf{0.0407$\pm$0.0219}& \textbf{0.1376$\pm$0.3660}\\
\bottomrule
\end{tabular}
\caption{Statistics of IC and correlations of mined alphas on CSI300.}
\label{tab:corr}
\end{table}

For a robust strategy, the alphas are expected to be diverse, having low Pearson Correlation Coefficients with each other. To answer the first question, we compute the correlations between alphas generated by gplearn, AlphaGen, and \name. The result is shown in Tab.~\ref{tab:corr}.

From the table, we can see that \name\ generates the best set of alphas in terms of IC. Meanwhile, it generates the most diverse set of alphas, as measured by the mean correlation. It is worth noting that gplearn generates a set of alphas with high correlations.  The high correlation results from the minor mutation of constants in the alpha expressions after it gets trapped in a local optima. With the more diverse and better alpha set, \name\ has greater potential to generate a more robust trading strategy.

\subsection{Performance of Generated Alphas}

\renewcommand{\arraystretch}{1.2} 
\begin{table}[htbp]
\vspace{-5mm}
\centering
\begin{tabular}{c|cccc}
\toprule
\multirow{2}{*}{\textbf{Method}} & \multicolumn{2}{c}{\textbf{CSI300}} & \multicolumn{2}{c}{\textbf{CSI500}} \\
\cline{2-5}
 & \textbf{IC} & \textbf{Rank IC} & \textbf{IC} & \textbf{Rank IC} \\
\hline
MLP & 0.0123$\pm$0.0006 & 0.0178$\pm$0.0017 & 0.0158$\pm$0.0014 &0.0211$\pm$0.0007 \\
XGBoost & 0.0192$\pm$0.0021& 0.0241$\pm$0.0027 & 0.0173$\pm$0.0017 & 0.0217$\pm$0.0022\\
LightGBM & 0.0158$\pm$0.0012 & 0.0235$\pm$0.0030 & 0.0112$\pm$0.0012  & 0.0212$\pm$0.0020 \\
gplearn & 0.0445$\pm$0.0044 & 0.0673$\pm$ 0.0058 & 0.0557$\pm$0.0117 & 0.0665$\pm$0.0154 \\
AlphaGen & 0.0500$\pm$0.0021 & 0.0540$\pm$0.0035 & 0.0544$\pm$0.0011& 0.0722$\pm$0.0017\\
\hline
\textbf{Ours} & \textbf{0.0576$\pm$0.0022} & \textbf{0.0681$\pm$0.0041} & \textbf{0.0612$\pm$0.0051} & \textbf{0.0731$\pm$0.0093}\\
\bottomrule
\end{tabular}
\caption{Performance on CSI300 and CSI500 in the test dataset.}
\label{tab:IC_perf}
\end{table}

To answer the second question, we run the baselines and our method on the CSI300 and CSI500 stock datasets and evaluate them on the two metrics. The results are shown in Tab.~\ref{tab:IC_perf}.

The first three methods, MLP, XGBoost, and LightGBM, combine an existing set of alphas from Qlib. They have a worse performance due to the usage of the open-source set of alphas. On the other hand, gplearn and AlphaGen are based on formulaic alphas generated by themselves. Alphas generated by gplearn and AlphaGen perform better on the test dataset. Although AlphaGen has designed a framework to filter alphas, it neither ensures the validity of alphas upon generation nor emphasizes diversity, which leads to possible performance degradation. We attribute the superior performance of \name\ to the logical soundness and diversity of alphas.

\subsection{Stock Market Backtest}
\begin{figure}[t]
    \centering
    \includegraphics[width=1.0\linewidth, trim=70 40 90 60,clip]{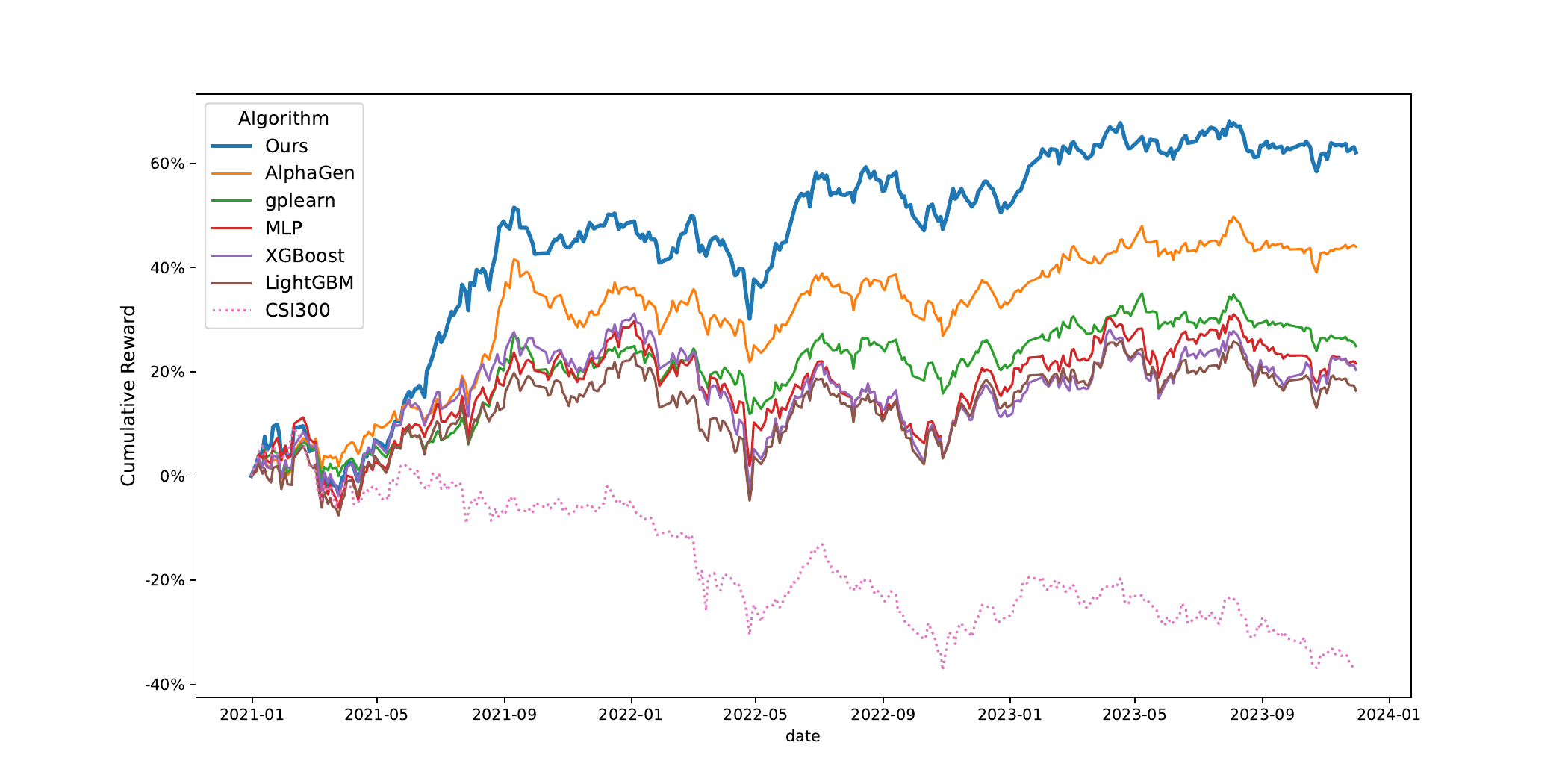}
    \caption{Backtest result on CSI300. The value of the y axis represents the cumulative reward.}
    \label{fig:backtest}
    \vspace{-2mm}
\end{figure}

To further validate the effectiveness of our method, we conduct an experiment on a simulated environment. The data is from the Chinese A-shares market in the test period. The trading strategy is a top-k/drop-n strategy. On each trading day, stocks are first sorted according to the alpha values, then the top-$k$ stocks are selected to trade. With at most $n$ stocks traded every day, we try to invest evenly across the $k$ stocks. In out experiment, $k=50$ and $n=5$. The result of the backtest is shown in Fig.~\ref{fig:backtest}. Our method demonstrates superior performance on the CSI300 stock market.

\section{Conclusion}
In this work, we introduce \name, a novel framework for the discovery of formulaic alphas. Using RL and MCTS, we harness the power of modern machine learning techniques to address the challenges of discovering powerful formulaic alphas in the vast search space.
\name\ formulates the alpha generation process as a program construction task, using RL-guided MCTS as the search algorithm. Our refined value estimation, performance evaluation and dimension check ensures the discovery of high-quality alphas and a good and robust trading strategy, which are validated in the experiments.
From the perspective of search algorithms, vast theoretical research has been done on MCTS and RL. Our method benefits from the existing research. The search algorithm can minimize the regret to the ground truth within theoretical bounds, compared to the mostly empirical results of previous methods.
On the engineering side, we propose a novel framework to generate formulaic alphas. This framework allows a general design of search space pruning for formulaic alphas, including but not limited to dimensional consistency rules. 
\newpage
\bibliography{main}

\begin{thebibliography}{15}
\providecommand{\natexlab}[1]{#1}
\providecommand{\url}[1]{\texttt{#1}}
\expandafter\ifx\csname urlstyle\endcsname\relax
  \providecommand{\doi}[1]{doi: #1}\else
  \providecommand{\doi}{doi: \begingroup \urlstyle{rm}\Url}\fi

\bibitem[Chen \& Guestrin(2016)Chen and Guestrin]{chen2016xgboost}
Tianqi Chen and Carlos Guestrin.
\newblock {XGB}oost: A scalable tree boosting system.
\newblock In \emph{Proceedings of the 22nd {ACM SIGKDD I}nternational
  {C}onference on {K}nowledge {D}iscovery and {D}ata {M}ining}, pp.\  785--794,
  2016.

\bibitem[Cui et~al.(2021)Cui, Wang, Zhang, Chen, Luo, and
  Ooi]{cui2021alphaevolve}
Can Cui, Wei Wang, Meihui Zhang, Gang Chen, Zhaojing Luo, and Beng~Chin Ooi.
\newblock Alpha{E}volve: A learning framework to discover novel alphas in
  quantitative investment.
\newblock In \emph{Proceedings of the 2021 {I}nternational {C}onference on
  {M}anagement of {D}ata}, pp.\  2208--2216, 2021.

\bibitem[Dam et~al.(2019)Dam, Klink, D'Eramo, Peters, and
  Pajarinen]{dam2019generalized}
Tuan Dam, Pascal Klink, Carlo D'Eramo, Jan Peters, and Joni Pajarinen.
\newblock Generalized mean estimation in {M}onte-{C}arlo tree search.
\newblock \emph{arXiv preprint arXiv:1911.00384}, 2019.

\bibitem[Kamienny et~al.(2023)Kamienny, Lample, Lamprier, and
  Virgolin]{kamienny2023deep}
Pierre-Alexandre Kamienny, Guillaume Lample, Sylvain Lamprier, and Marco
  Virgolin.
\newblock Deep generative symbolic regression with {M}onte-{C}arlo-tree-search.
\newblock In \emph{International Conference on Machine Learning}, pp.\
  15655--15668, 2023.

\bibitem[Ke et~al.(2017)Ke, Meng, Finley, Wang, Chen, Ma, Ye, and
  Liu]{ke2017lightgbm}
Guolin Ke, Qi~Meng, Thomas Finley, Taifeng Wang, Wei Chen, Weidong Ma, Qiwei
  Ye, and Tie-Yan Liu.
\newblock Light{GBM}: A highly efficient gradient boosting decision tree.
\newblock In \emph{Advances in Neural Information Processing Systems}, pp.\
  3146--3154, 2017.

\bibitem[Koza(1994)]{koza1994genetic}
John~R Koza.
\newblock Genetic {P}rogramming as a means for programming computers by natural
  selection.
\newblock \emph{Statistics and {C}omputing}, 4:\penalty0 87--112, 1994.

\bibitem[Mankowitz et~al.(2023)Mankowitz, Michi, Zhernov, Gelmi, Selvi,
  Paduraru, Leurent, Iqbal, Lespiau, Ahern, K{\"{o}}ppe, Millikin, Gaffney,
  Elster, Broshear, Gamble, Milan, Tung, Hwang, Cemgil, Barekatain, Li,
  Mandhane, Hubert, Schrittwieser, Hassabis, Kohli, Riedmiller, Vinyals, and
  Silver]{alphadev}
Daniel~J. Mankowitz, Andrea Michi, Anton Zhernov, Marco Gelmi, Marco Selvi,
  Cosmin Paduraru, Edouard Leurent, Shariq Iqbal, Jean{-}Baptiste Lespiau, Alex
  Ahern, Thomas K{\"{o}}ppe, Kevin Millikin, Stephen Gaffney, Sophie Elster,
  Jackson Broshear, Chris Gamble, Kieran Milan, Robert Tung, Minjae Hwang,
  A.~Taylan Cemgil, Mohammadamin Barekatain, Yujia Li, Amol Mandhane, Thomas
  Hubert, Julian Schrittwieser, Demis Hassabis, Pushmeet Kohli, Martin~A.
  Riedmiller, Oriol Vinyals, and David Silver.
\newblock Faster sorting algorithms discovered using deep reinforcement
  learning.
\newblock \emph{Nature}, 618\penalty0 (7964):\penalty0 257--263, 2023.

\bibitem[Mundhenk et~al.(2021)Mundhenk, Landajuela, Glatt, Santiago, Faissol,
  and Petersen]{mundhenk2021symbolic}
T~Nathan Mundhenk, Mikel Landajuela, Ruben Glatt, Claudio~P Santiago, Daniel~M
  Faissol, and Brenden~K Petersen.
\newblock Symbolic regression via neural-guided genetic programming population
  seeding.
\newblock \emph{arXiv preprint arXiv:2111.00053}, 2021.

\bibitem[Sahoo et~al.(2018)Sahoo, Lampert, and Martius]{sahoo2018learning}
Subham Sahoo, Christoph Lampert, and Georg Martius.
\newblock Learning equations for extrapolation and control.
\newblock In \emph{International Conference on Machine Learning}, pp.\
  4442--4450, 2018.

\bibitem[Silver et~al.(2016)Silver, Huang, Maddison, Guez, Sifre, van~den
  Driessche, Schrittwieser, Antonoglou, Panneershelvam, Lanctot, Dieleman,
  Grewe, Nham, Kalchbrenner, Sutskever, Lillicrap, Leach, Kavukcuoglu, Graepel,
  and Hassabis]{alphazero}
David Silver, Aja Huang, Chris~J. Maddison, Arthur Guez, Laurent Sifre, George
  van~den Driessche, Julian Schrittwieser, Ioannis Antonoglou, Vedavyas
  Panneershelvam, Marc Lanctot, Sander Dieleman, Dominik Grewe, John Nham, Nal
  Kalchbrenner, Ilya Sutskever, Timothy~P. Lillicrap, Madeleine Leach, Koray
  Kavukcuoglu, Thore Graepel, and Demis Hassabis.
\newblock Mastering the game of {G}o with deep neural networks and tree search.
\newblock \emph{Nature}, 529\penalty0 (7587):\penalty0 484--489, 2016.

\bibitem[Silver et~al.(2017)Silver, Schrittwieser, Simonyan, Antonoglou, Huang,
  Guez, Hubert, Baker, Lai, Bolton, Chen, Lillicrap, Hui, Sifre, van~den
  Driessche, Graepel, and Hassabis]{alphago}
David Silver, Julian Schrittwieser, Karen Simonyan, Ioannis Antonoglou, Aja
  Huang, Arthur Guez, Thomas Hubert, Lucas Baker, Matthew Lai, Adrian Bolton,
  Yutian Chen, Timothy~P. Lillicrap, Fan Hui, Laurent Sifre, George van~den
  Driessche, Thore Graepel, and Demis Hassabis.
\newblock Mastering the game of go without human knowledge.
\newblock \emph{Nature}, 550\penalty0 (7676):\penalty0 354--359, 2017.

\bibitem[Tulchinsky(2019)]{Tulchinsky2015FindingAA}
Igor Tulchinsky.
\newblock Finding alphas: A quantitative approach to building trading
  strategies.
\newblock John Wiley \& Sons, 2019.

\bibitem[Yang et~al.(2020)Yang, Liu, Zhou, Bian, and Liu]{yang2020qlib}
Xiao Yang, Weiqing Liu, Dong Zhou, Jiang Bian, and Tie-Yan Liu.
\newblock Qlib: An {AI}-oriented quantitative investment platform.
\newblock \emph{arXiv preprint arXiv:2009.11189}, 2020.

\bibitem[Yu et~al.(2023)Yu, Xue, Ao, Pan, He, Tu, and He]{AlphaGen}
Shuo Yu, Hongyan Xue, Xiang Ao, Feiyang Pan, Jia He, Dandan Tu, and Qing He.
\newblock Generating synergistic formulaic alpha collections via reinforcement
  learning.
\newblock In \emph{Proceedings of the 29th {ACM SIGKDD C}onference on
  {K}nowledge {D}iscovery and {D}ata {M}ining}, pp.\  5476--5486, 2023.

\bibitem[Zhang et~al.(2020)Zhang, Li, Jin, and Li]{zhang2020autoalpha}
Tianping Zhang, Yuanqi Li, Yifei Jin, and Jian Li.
\newblock Auto{A}lpha: An efficient hierarchical evolutionary algorithm for
  mining alpha factors in quantitative investment.
\newblock \emph{arXiv preprint arXiv:2002.08245}, 2020.

\end{thebibliography}
\bibliographystyle{paper}
\end{document}